\documentclass[twocolumn,pra,aps,superscriptaddress]{revtex4-1}
\usepackage[english]{babel}
\usepackage[utf8]{inputenc}
\usepackage{url,psfrag,graphicx}
\usepackage{amsmath,amssymb,amsthm}
\usepackage{bm,xcolor}
\usepackage{hyperref}
\usepackage{physics}
\def\({\left(}
\def\){\right)}
\def\C{\mathbb{C}}
\def\Z{\mathbb{Z}}
\def\<{\langle}
\def\>{\rangle}
\def\even{\textrm{even}}
\def\odd{\textrm{odd}}
\def\opt{\textrm{opt}}

\begin{document}

\title{Optimizing the spatial spread of a quantum walk}

\author{Gonzalo Martín-Vázquez}
\email{gonmarvaz@gmail.com}
\affiliation{Departamento de F\'isica Te\'orica, Universidad Complutense de Madrid, Plaza de Ciencias 1, 28040 Madrid, Spain} 
\affiliation{Facultad de Ciencias Experimentales, Universidad Francisco de Vitoria, Carretera Pozuelo-Majadahonda km. 1,800, 28223 Pozuelo de Alarcón, Madrid, Spain}  

\author{Javier Rodríguez-Laguna}
\affiliation{Dpto. de Física Fundamental, Universidad Nacional
  de Educación a Distancia (UNED), Madrid, Spain}

\date{April 10, 2020}

\begin{abstract}
  We devise a protocol to build 1D time-dependent quantum walks in 1D
  maximizing the spatial spread throughout the procedure. We allow
  only one of the physical parameters of the coin-tossing operator to
  vary, i.e. the angle $\theta$, such that for $\theta=0$ we have the
  $\hat\sigma_z$, while for $\theta=\pi/4$ we obtain the Hadamard
  gate. The optimal $\theta$ sequences present non-trivial patterns,
  with mostly $\theta\approx 0$ alternated with $\theta\approx \pi/4$
  values after increasingly long periods. We provide an analysis of
  the entanglement properties, quasi-energy spectrum and survival
  probability, providing a full physical picture.
\end{abstract}

\maketitle


\section{Introduction}
\label{sec:intro}

Quantum walks (QW) are the quantum analogues of classical random walks
(CRW). First suggested by Feynman and Hibbs in 1965 \cite{feynman65},
quantum walks were described by Aharonov {\em et al.} in 1993
\cite{aharonov93}, where it was noted that they give rise to a more
intrincate probability distribution due to quantum
interference. Moreover, quantum walks may spread much faster than
their classical counterparts. Indeed, the spatial deviation of a
classical random walk grows diffusively with time ($\sigma \propto
t^{1/2}$), while it can be ballistic for a quantum walk ($\sigma
\propto t$).

Similarly to the classical case, there are two main types of quantum
walks: continuous-time quantum walks (CTQWs) and discrete-time quantum
walks (DTQWs), which will be the focus of this work. Positions are
usually discrete in DTQW (yet, see \cite{mlodinov18}). In CTQW,
evolution is ruled by a Schrödinger equation, while in a DTQW the
system is endowed with an internal degree of freedom (coin space) and
a configuration space (position space) representing the walker's
position. The system evolves in discrete time steps by applying a
certain coin-toss operator on the coin space and a conditional
displacement in the position space \cite{venegas12}. DTQW have been
succesfully implemented experimentally in different setups: nuclear
magnetic resonance (NMR) \cite{ryan05}, waveguide arrays
\cite{perets08, sansoni12}, ion traps \cite{schmitz09} and
superconducting circuits \cite{flurin17}.

Quantum walks present a rich range of behaviors upon changing their
parameters or introducing decoherence in the system. In presence of
dynamical disorder (time-dependent random parameters) and/or quenched
disorder (position dependent), the time evolution of a DTQW can change
completely, approaching a Gaussian-like distribution in position
space, similarly to the classical situation. Dynamical disorder leads
the system to develop maximal entanglement between the coin and the
positional degrees of freedom, while in the case of quenched disorder
we can observe Anderson localization \cite{vieira14, gullo17, zeng17,
  liu18}. Recently, this effect has been demonstrated experimentally
\cite{wang18}, and the idea of searching for an optimum sequence to
maximize the entanglement has been suggested. Moreover, it was found
in \cite{orthey18} that the fastest route to entangle the system up to
its maximum value is to alternate between ordered and disordered
parameters. There is also an interesting interplay between
localization-delocalization transitions depending on the statistical
regime of the randomness \cite{mendes19} and the lack of periodicity
of the spatial inhomogeneties \cite{buarque19}, suggesting a very rich
dynamics.
 
One of the most promising uses of the quantum walk is the development
of novel quantum algorithms \cite{ambainis03}. Interestingly, it has
been demonstrated that quantum walks can perform universal quantum
computation both for CTQW \cite{childs09} and DTQW
\cite{lovett10}. Classical random walks have been used for simulated
annealing purposes for various decades \cite{kirkpatrick83}. Their
quantum counterparts might benefit both from a faster spread rate and
from interference effects. A CTQW-based algorithm has been proposed
presenting an exponential speed-up to traverse a special type of
graph, called the {\em glued-trees problem} \cite{childs03}, while
DTQW can be used to implement Grover's algorithm in order to search in
an unstructured database \cite{shenvi03, lovett19}, achieving a
quadratic speed-up.

Uniform spread of a quantum walk can help sample a large problem space
\cite{kendon03, maloyer07}. Moreover, it could be useful for
initializing a system in an unbiased state for searching problems
\cite{callison19} or to determine its statistical properties
\cite{orthey17, ghizoni19}. Decoherence (or, alternatively,
measurement) can optimize the spreading and mixing properties of a
quantum walk \cite{kendon03}, improving its computational properties
\cite{maloyer07}. However, decoherence reduces the spreading rate in
the long run, becoming diffusive as in the classical case
\cite{brun03, brun03v2}. Interestingly, for short running times $T$, a
certain amount of decoherence can make the distribution very close to
a uniform one, retaining the ballistic spreading \cite{kendon03}. Yet,
the spatial spread grows as $T/\sqrt{2}$ instead of the maximum
possible value $T$. It has also been shown that decoherence in
position space (introduced as a noise that can shift positions) gives
rise to a smooth probability distribution while mantaining the quantum
properties, such as the ballistic propagation and the entanglement
between the coin and the position \cite{annabestani16}.

In this work we show that a nearly uniform spatial distribution can be
obtained for all times, with maximum ballistic spread and without
decoherence. The procedure involves the use of a time-dependent
coin-tossing unitary operator. As we will show, the time-dependent
protocol is stable, i.e. it admits small perturbations maintaining the
spatial properties.

The idea of a uniform distribution in position space could be also of
interest in biology, specifically in the analysis of the light
harvesting processes, such as photosynthesis
\cite{arndt09}. Experimental work has found that the process depends
on the delocalization of the exciton over the molecules
\cite{brixner05, engel07, panitchayangkoon10}, and it has been
proposed that its high efficiency could be explained by means of a
quantum search algorithm \cite{engel07}, specifically, one based on a
quantum walk \cite{mohseni08}. Indeed, time-dependent quantum walks
providing uniform sampling of the search space might provide an
interesting advantage.

This paper is organized as follows. The model is introduced in
Sec. \ref{sec:model}, along with our target function describing the
spatial spread of the quantum walker. Sec. \ref{sec:results} exposes
the numerical results, with special emphasis on the characterization
of the optimal set of operators. The discussion of the physical meaning
of our results is performed in Sec. \ref{sec:understanding}, employing
the spectral properties of the optimal evolution operator and the
analytical properties of the survival
probability. Sec. \ref{sec:conclusions} is devoted to our conclusions
and suggestions for further work.


\section{Spatial spread of a quantum walk}
\label{sec:model}

Let us consider a discrete-time quantum walker (DTQW), consisting of a
particle moving on an infinite 1D chain, known as {\em position
  space}, endowed with an internal degree of freedom, known as {\em
  coin space}. Position space is spanned by the basis vectors
$\ket{x}_p$ with $x\in\Z$, and the coin space is just $\C^2$, spanned
by states $\ket{L}_c$ and $\ket{R}_c$. Thus, the system state is
spanned by tensor product states of particle and coin,
$\ket{x,c}=\ket{x}_p\otimes \ket{c}_c$, with $x\in\Z$ and $c\in
\{L,R\}$. Thus, the total wavefunction can be always expressed as

\begin{equation}
  \ket{\psi}=\sum_{x,c} \psi_{x,c} \ket{x,c},
  \label{eq:psi}
\end{equation}
Thus, the probability that the walker will be found at position $x$
will be given by

\begin{equation}
  P_x=|\psi_{x,L}|^2+|\psi_{x,R}|^2.
  \label{eq:pn}
\end{equation}

\begin{figure*}
\includegraphics[width=1\textwidth]{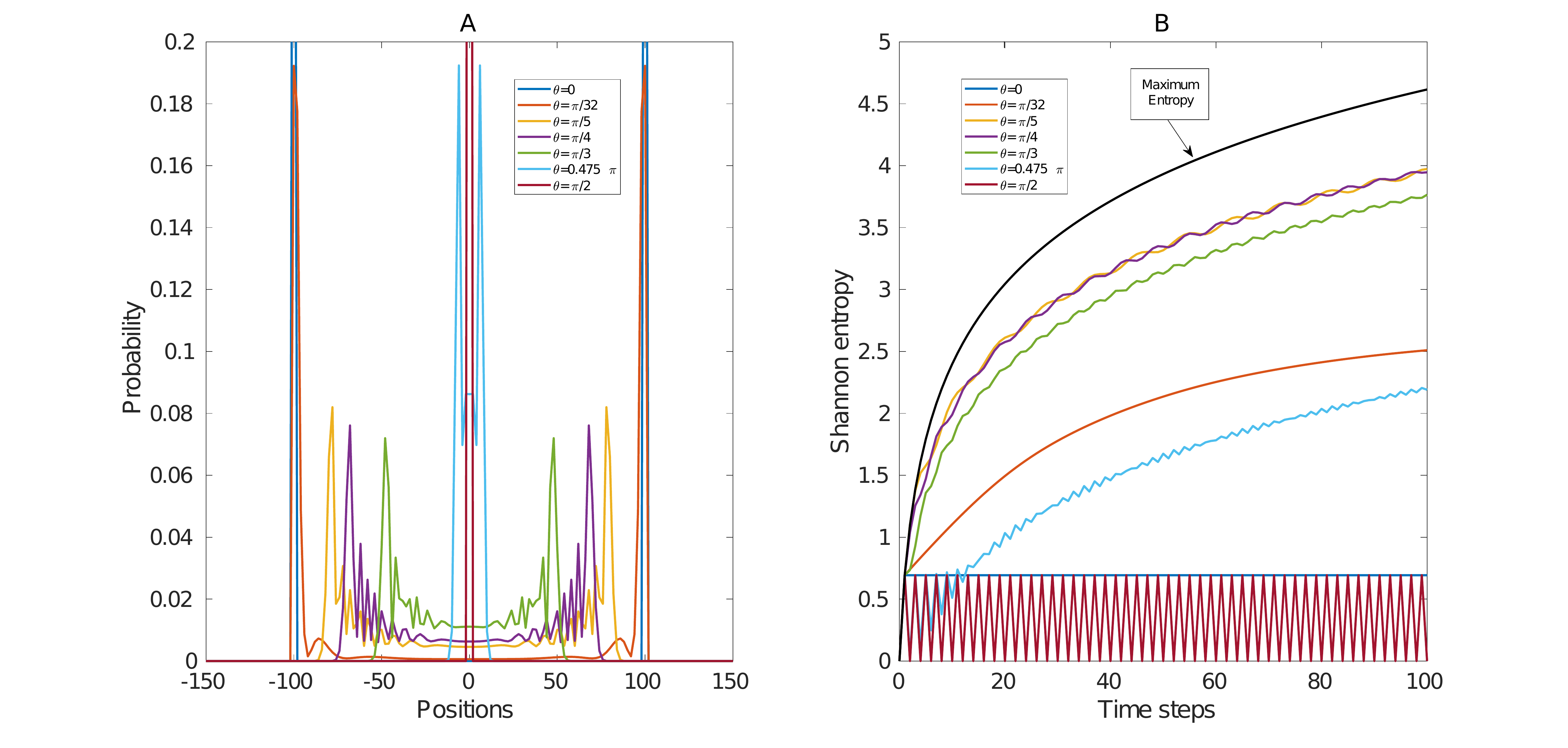}
\caption{(A) Probability distribution in position space for a
  discrete-time quantum walker on a line after $T=100$ time
  steps. Each curve is characterized by a constant value of the
  parameter $\theta$, while $\xi=\zeta=\pi/2$. (B) Time-evolution of
  the Shannon entropy, Eq. \eqref{eq:shannon}, for the same cases
  shown in A. The values of $\theta\in
  \{0,\pi/32,\pi/5,\pi/4,\pi/3,19\pi/40,\pi/2\}$ are color coded. }
\label{fig1}
\end{figure*}

The time evolution of the system its obtained through the consecutive
application of unitary operators, each of them consisting of a coin
tossing unitary operator and and conditional shift in position
space. The coin operator can be written as a SU(2) matrix

\begin{equation}
\hat{B}(\xi,\theta,\zeta)=
  \begin{pmatrix}
    e^{i\xi}\cos\theta & e^{i\zeta}\sin\theta \\
    -e^{-i\zeta}\sin\theta & e^{-i\xi}\cos\theta 
  \end{pmatrix},
\label{coinop}
\end{equation}
where $\theta\in[0,\frac{\pi}{2}]$ and $\xi,\zeta\in[0,2\pi]$
\cite{chandrashekar08, chandrashekar09}. Setting
$\xi=\zeta=\frac{\pi}{2}$ we get

\begin{equation}
\hat{B}(\theta)=
  \begin{pmatrix}
    \cos\theta & \sin\theta \\
    \sin\theta & -\cos\theta 
  \end{pmatrix},
\label{coinop2}
\end{equation}
up to a global phase. Note that \eqref{coinop2} reduces to the usual
Hadamard operator when $\theta=\frac{\pi}{4}$.

The shift operator yields the displacement of the particle in position
space conditioned by the internal degree of freedom of the coin, and
can be written as

\begin{equation}
  \hat{S}=\sum_{x=-\infty}^{\infty}
  \(\ket{x+1}\bra{x}\otimes\ket{R}\bra{R}
  +\ket{x-1}\bra{x}\otimes\ket{L}\bra{L}\).
\label{shift}
\end{equation}
In practice, we will consider a finite-dimensional version of
Eq. \eqref{shift}, with specific boundary conditions (see Appendix
\ref{sec:appendix_A}). Finally, the total unitary evolution operator
is given by

\begin{equation}
\
\hat{U}(\xi,\theta,\zeta)=\hat{S}\cdot(\hat{\mathbb{I}}_p \otimes
\hat{B}(\xi,\theta,\zeta)),
\label{unitary}
\end{equation}
such that

\begin{equation}
\ket{\psi(t+1)}=\hat{U}(\xi,\theta,\zeta)\ket{\psi(t)}.
\label{evolution}
\end{equation}
Since we will consider time-dependent parameters in \eqref{coinop} and
\eqref{coinop2}, the evolution of the system for $T$ time steps will
be given by

\begin{equation}
\ket{\psi(t+T)}=\hat{U}_{T}\cdots\hat{U}_1\ket{\psi(t)},
\label{evolutiontime}
\end{equation}
where $\hat{U}_t=\hat{U}(\xi_t,\theta_t,\zeta_t)$ for time $t$.
As our initial state, we will consider a particle localized at
$x=0$ and with a coin component of the form

\begin{equation}
  \ket{\psi_S} \equiv \ket{\psi(0)} =
  \frac{1}{\sqrt{2}} \ket{0}_p\otimes\(\ket{R}_c+i\ket{L}_c\),
\label{initial}
\end{equation}
leading to a left-right symmetric evolution for $\xi=\zeta=\pi/2$
\cite{kendon03}. In this work we will only consider quantum walkers
characterized by a sequence $\{\theta_t\}_{t=1}^T$, with
$\xi_t=\zeta_t=\pi/2$ for all time.

As we can see in Fig. \ref{fig1}A, the typical behavior of the quantum
walk using a constant coin-tossing operator is far from being uniform
in position space. Instead, the probability distributions show an
intrincate interference pattern. The maximal spread is obtained for
$\theta=0$, corresponding to $\hat{B}(0)\equiv \hat{\sigma}_z$, and
the minimal one is found for $\theta=\pi/2$, which corresponds to
$\hat{B}(\pi/2)\equiv \hat{\sigma}_x$.

The spread of the probability distribution in position space can be
characterized using Shannon's entropy,

\begin{equation}
  S=-\sum_x P_x \log P_x,
  \label{eq:shannon}
\end{equation}
with $P_x$ given in Eq. \eqref{eq:pn}. After $t$ time-steps, the
maximal value possible for the entropy is given by
$S_{\text{max}}(t)=\log(t+1)$. This bound can be understood by
noticing that, after $t$ time-steps, the particle can only reach
$2t+1$ sites, but only odd (even) positions can be occupied after an
odd (even) number of time-steps, in absence of
decoherence. Fig. \ref{fig1}B shows the time-evolution of the
Shannon's entropy of a quantum walker for different constant values of
$\theta$. Indeed, the maximal bound is never reached, yet for some
values of $\theta$ we obtain a logarithmic growth, corresponding to a
ballistic spread.


\subsection{Optimizing quantum walks}

The aim of this work is to obtain the optimal sequence of coin-tossing
operators maximizing the spatial spread of the quantum walker along
its whole history, up to a certain time-step $T$. We will restrict our
search to discrete-time quantum walkers without decoherence and with
coin-tossing operators using $\xi_t=\zeta_t=\pi/2$, i.e.: they will be
fully determined by the sequence $\{\theta_t\}_{t=1}^T$.

For a fixed time-step $t$, a good figure of merit is given by the
Shannon entropy of the spatial probability distribution,
Eqs. \eqref{eq:shannon} and \eqref{eq:pn}, normalized by the
maximal value achievable for that time-step. After $t$ time-steps, the
walker can reach a total of $t+1$ sites (not $2t+1$ as one might
naively expect, because the walker can only reach even-indexed sites
after an even number of steps, and viceversa). Thus, the maximal
achievable Shannon entropy after $t$ time-steps is
$S_{\text{max}}=\log(t+1)$. Therefore, a reasonable observable to
characterize the extent of the spread of the quantum walker after $T$
time-steps is given by

\begin{equation}
  F\(\theta_0,\cdots,\theta_T\)
  =1-{1\over T+1}
  \sum_{t=0}^T
    \frac{S(t)}{\log(t+1)},
\label{F}
\end{equation}
where $S(t)$ is the Shannon entropy after $t$ time-steps, given in
Eq. \eqref{eq:shannon}. This magnitude $F$ reaches its maximum value
$F=1$ when the walker is completely localized, while its minimum $F=0$
corresponds to our desired situation, when the spread is maximal along
its whole history.

Finding the optimal set of $\{\theta_t\}$ which minimizes $F$ is a
computationally demanding task. We employ a combination of conjugated
gradients method and sampling of initial configurations in order to
achieve the global minimum when the target function presents many
local minima, as it has been done by other authors
\cite{santos18}. The number of initial configurations employed was 50
for moderate times, and as high as 200 for the maximal time reached,
$T=45$. Our numerical experiments allow us to conjecture that the
optimization landscape is rather complex, as it will be discussed in
the next section.


\section{Results}
\label{sec:results}

\subsection{Different approaches to optimize spread}

Our first attempt at obtaining the optimal set of parameters
$\{\theta_t\}_{t=1}^T$ in \eqref{coinop2} for an optimal spread is
analytical. For $T<4$, we have found the optimal distribution
corresponding to the maximal spread, the detailed calculations are
provided in Appendix \ref{sec:appendix_B}, here we will only cite the
main results. First of all, notice that the spread does not depend on
$\theta_1$, so this first value is always arbitrary. For the second
and third steps, we obtain $\theta_2=\arctan(1/\sqrt{2})$ and
$\theta_3=\frac{\pi}{6}$, respectively. For $T\geq 4$, we have proved
that no set of coin-tossing operators will yield this perfect
spread. Yet, a numerical evaluation of the $\theta$ sequences yielding
an optimal amount of spread is still possible, and the following
section is devoted to their characterization.

\begin{figure*}
\includegraphics[width=1\textwidth]{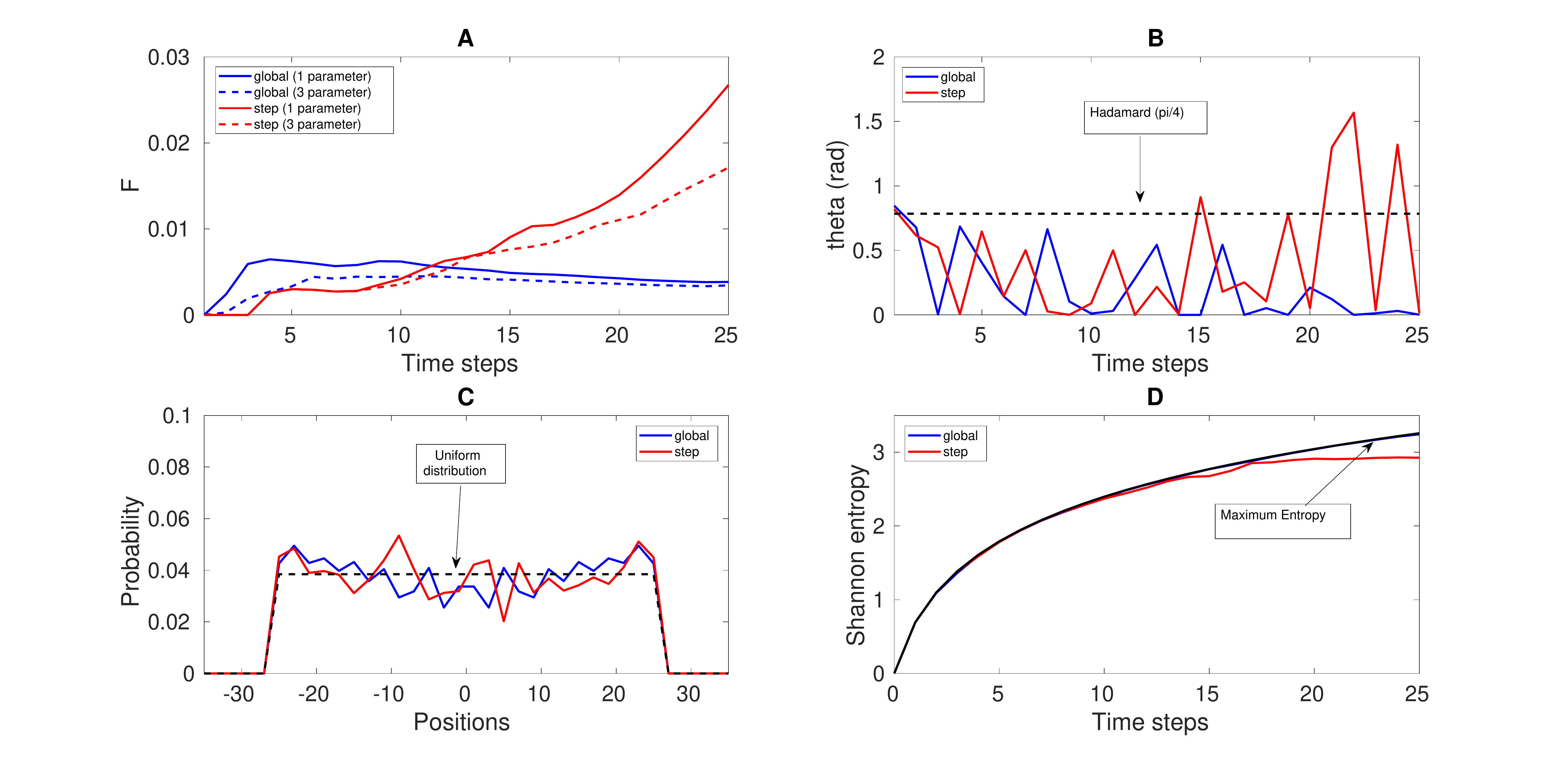}
\caption{\label{fig2} Optimization of the spread, obtained minimizing
  the value of $F$. We employ two different approaches: global or
  step-by-step. Moreover, we explore the use of a single coin
  parameter ($\theta$) or the full set of three parameters. The
  results are shown up to $T=25$ time steps. (A) Evolution of the
  minimal $F$ value, as obtained using the different optimization
  criteria. Note that the actual (and computed) value for the global
  case is $T=25$; the previous time steps are reconstructed {\em a
    posteriori}. (B) Sequence of $\theta$ parameters that minimize
  $F$. (C) Probability distribution in position space. (D) Shannon
  entropy of the probability distribution in position space. Note that
  the global case overlaps with the maximum entropy case.}
\end{figure*}

For a given number $T$ of time steps, there are a few different
approaches to the optimization of the set of $\theta$ parameters.

\begin{itemize}
\item We may minimize a single {\em global} value of $F$ spanning
  time-steps 1 to $T$, i.e. obtain the whole set of $\theta$
  parameters in a single optimization procedure.

\item Alternatively, we can operate through a {\em step-by-step}
  minimization: once the sequence $\theta_1$ to $\theta_t$ is
  optimized, we obtain the optimal value of $\theta_{t+1}$, and
  iterate up to $t=T$.

\item Finally, we can optimize only the value of the final spread,
  after $T$ time-steps, disregarding the intermediate stages. This
  procedure always results in an optimal spread. We will not discuss
  this approach further in the main text, and leave the details for
  Appendix \ref{sec:appendix_C}.
\end{itemize}

In Fig. \ref{fig2} we compare the first two approaches. The
step-by-step method achieves a better optimization for short times,
but the value of $F$ (which measures our failure to obtain perfect
spread) increases fast between $T=10$ and $T=15$, and for large times
the global procedure is considerably better (Fig. \ref{fig2}A). Note
that the global approach only provides a single value of $F$,
corresponding to the final time step; but in the figure we provide an
{\em a posteriori reconstruction} of the $F$ values for all times. The
sequence of $\theta$ parameters is different for each approach, with
some unexpected differences. For example, all $\theta$ values are
below $\frac{\pi}{4}$ for the global approach, while they can reach
values above that threshold for the step-by-step procedure,
specifically near the time where the technique starts to fail
(Fig. \ref{fig2}B). We can see the final probability distribution for
the two approaches and its difference with a perfectly uniform one in
Fig. \ref{fig2}C. Finally, in Fig. \ref{fig2}D we can observe the
evolution of the Shannon entropy of the probability distribution,
compared to its maximal possible value. Notice how the globally
obtained entropy remains close this maximum possible value, while the
step-by-step entropy deviates from it. Henceforth, given its higher
precision, we will make use of the global approach in the rest of
this work.

The minimal value of $F$ obtained using all three parameters of the
coin-tossing operator ($\xi$, $\theta$ and $\zeta$ in
Eq. \eqref{coinop}) will be always equal or lower than the value
obtained using only the parameter $\theta$ and $\xi=\zeta=\pi/2$
(i.e., using Eq. \eqref{coinop2}). Interestingly, the difference gets
smaller with time when we follow the global optimization approach, as
we can see in Fig. \ref{fig2}A. Similar results have been reported
when analyzing entanglement properties \cite{vieira14}. Henceforth we
will consider only the coin operator \eqref{coinop2} with one
parameter.

\subsection{Characterization of the Optimal Sequences}

\begin{figure*}
\includegraphics[width=20cm]{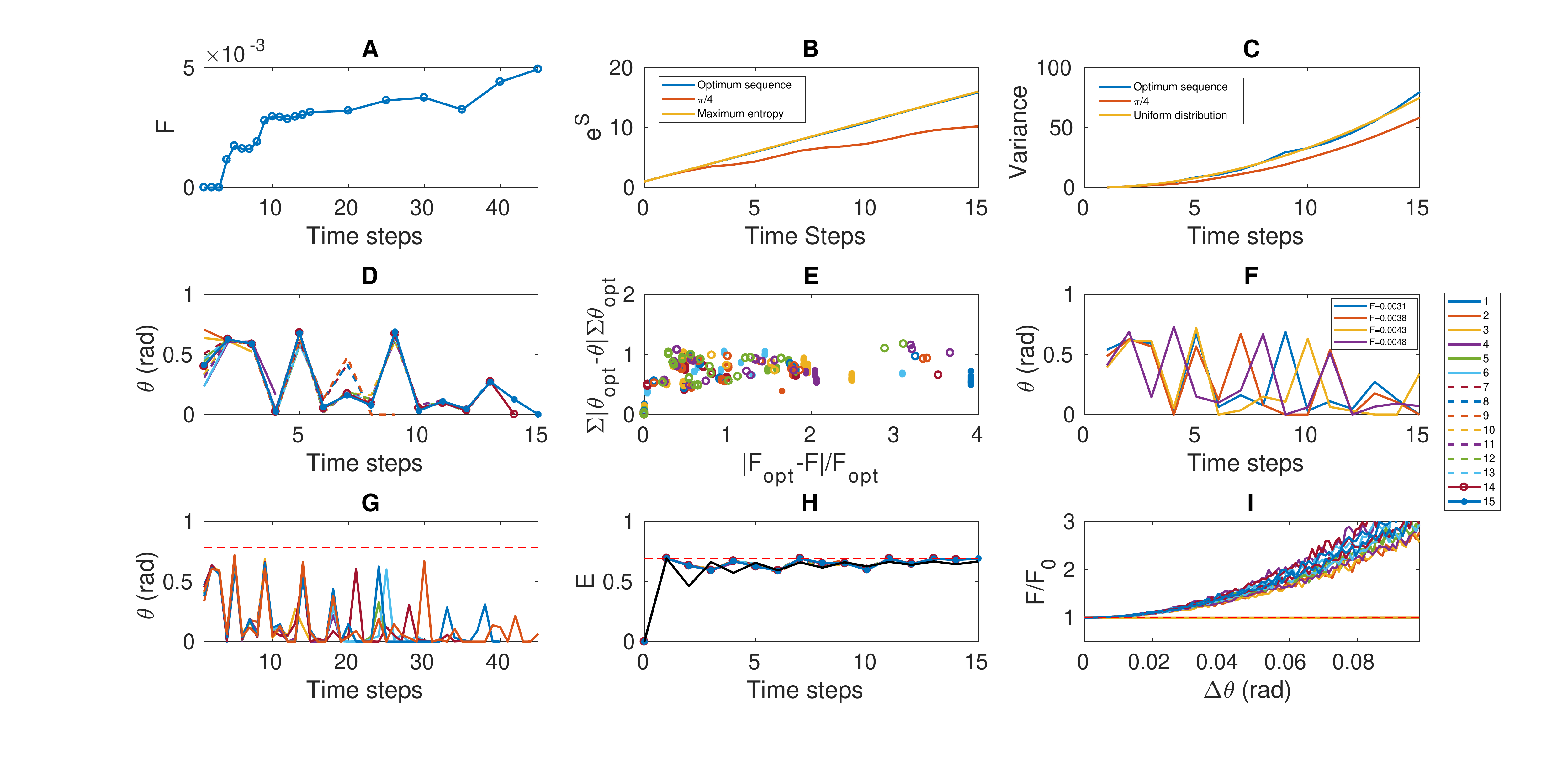}
\caption{\label{fig3} Optimized sequences of the $\theta$ parameters
  for $T=1,\cdots,15$. (A) Minimal $F$ values for different $T$ under
  global optimization. (B) Exponential of the Shannon entropy ($S$) of
  the probability distribution in position space and (C) variance for
  the optimum sequence for $T=15$, compared to the results using the
  Hadamard coin operator ($\theta=\frac{\pi}{4}$). Note that the
  Shannon entropy for the optimum case overlaps with the maximum
  entropy situation. (D) Optimized $\theta$ parameters for $T=1$ up to
  $T=15$. The horizontal dashed red line corresponds to
  $\frac{\pi}{4}$ (Hadamard). (E) Differences of local minima with
  respect to the optimum case computed as the difference in the $F$
  value and the $\theta$ parameter. Each symbol corresponds to
  optimization for different time steps ($T=1..15$). (F) Four lowest
  local minima for $T=15$. (G) Optimized sequences of $\theta$
  parameters for $T=5,10,15,20,25,30,35,40,45$. The horizontal dashed
  red line corresponds to $\frac{\pi}{4}$ (Hadamard). (H) Time
  evolution of the von Neumman entropy of the reduced density matrix
  as a measure of entanglement for the optimized sequences of $\theta$
  parameters for $T=1$ up to $T=15$. The black line corresponds to the
  mean value of 1000 simulations with random $\theta$ parameters. The
  horizontal red dashed line corresponds to the situation of maximum
  entanglement $\log(2)$. (I) Stability of the optimized sequences of
  $\theta$ parameters against increasing perturbations (noise) of the
  $\theta$ value (see text). For the case $T=1$ there is no change;
  and for $T=2,3$ the changes are not appreciable.}
\end{figure*}

Figure \ref{fig3}A shows how the minimal $F$ values increase as the
final time $T$ increases. As it was expected, the spread is perfect
for $T\leq 3$. As shown before, the Shannon entropy for the optimum
sequence ($T=15$) is very close to the maximum value when compared to
the usual Hadamard case (Fig. \ref{fig3}B). Let us remind the reader
that the variance in position space is defined by
$\sigma^2=\<x^2(t)\>-\<x(t)\>^2$. Thus, considering that for even
(odd) time steps only the even (odd) sites are occupied and that the
probability distribution in position space is uniform, the variance of
an idealized uniform quantum walk takes the form

\begin{equation}
\begin{split}
  \sigma_T^2 &= \frac{2}{T_\even+1} \sum_0^{T_\even/2} (2T_\even)^2 \\
 &= \frac{2}{T_\odd+1} \sum_0^{(T_\odd-1)/2} (2T_\odd+1)^2 \\
 &= \frac{1}{3} \frac{T(T^2+3T+2)}{T+1}.
\end{split}
\label{var}
\end{equation}
The evolution of the variance for the optimal sequence for $T=15$ is,
indeed, very similar to our analytical expression \eqref{var}, as we
can check in Fig. \ref{fig3}C. The optimal sequences of $\theta$
parameters are depicted in Fig. \ref{fig3}D. Unfortunately, they do
not present a regular pattern which can help us predict their
evolution for larger time spans. Yet, optimal sequences obtained for
low values of $T$ are very similar among themselves, but differences
become significant for optimal sequences corresponding to longer
times, as we can check in Fig. \ref{fig3}G. Nonetheless, there are
some manifest patterns in the optimal sequences, such as an
alternation between values close to $\theta=\frac{\pi}{4}$ and
$\theta=0$, with increasing periods. We will discuss this pattern
later in this section. Notice that all $\theta$ parameters are always
below $\theta_c=\frac{\pi}{4}$.

During the optimization procedure we obtain on occasion local minima
of the target function $F$ which do not correspond to the global
minimum, $F_\opt$. In Fig. \ref{fig3}E we have considered these local
minima. For each local solution we provide a point in the plot, where
the abscissa is given by $|F-F_\opt|$ and the ordinate provides the
difference in the $\theta$ values, defined as

\begin{equation}
  \Delta[\{\theta_i\}]\equiv {\sum_{i=1}^T
    |\theta_i^\opt-\theta_i|\over (\sum_{i=1}^T \theta_i^\opt)}.
  \label{eq:distheta}
\end{equation}
The resulting plot provides the image of a complex landscape, with a
great variety of local minima, typical of glassy systems, which might
be related to replica symmetry breaking \cite{ueda14, itoi17}. As an
illustration, Fig. \ref{fig3}F depicts the optimal sequence for
$T=15$ along with the three lowest-$F$ local minima.

Since we are neglecting decoherence, the complete system state
(particle and coin) remains pure throughout time evolution. Thus, we
can make use of the von Neumann entropy of the reduced density
matrices as a measure of the entanglement between particle and coin,

\begin{equation}
E(t)=-\Tr[\rho_1(t) \ln\rho_1(t)]=-\Tr[\rho_2(t) \ln\rho_2(t)],
\label{neu}
\end{equation}
where $\rho_{1,2}(t)=Tr_{2,1}\rho(t)$ are the reduced density matrices
of the position and coin degrees of freedom, respectively, and
$\rho(t)=\ket{\psi(t)}\bra{\psi(t)}$. We compare the entanglement of
the optimized sequences for different time steps with the the case of
random evolution of the $\theta$ parameters in Fig. \ref{fig3}H. As we
can readily see, the entanglement of the optimized sequences tends to
its maximum value, as in the random case, but slightly faster.

In order to test the robustness of the optimized $\theta$ sequences,
we have introduced an increasing amount of noise in the parameters,
$\theta_i \to \theta_i + \Delta\theta \cdot \eta_i$, where the
$\eta_i$ are i.i.d. Gaussian random variables of zero average and unit
variance. Let $F_0$ be the optimal value for $F$ for the maximal
$T$. For all values of the noise amplitude, $\Delta\theta$, we
evaluate $F$ for $N_s=10^3$ different random perturbations of the
optimal sequence, and the quotient $F/F_0$ is plotted in
Fig. \ref{fig3}I. For consistency, values of $\theta_i$ that leave the
range $[0,\pi/2]$ are automatically set to the closest extreme of the
interval. As expected, we observe a smooth increase of the optimal
value of $F$.

\begin{figure*}
\includegraphics[width=1\textwidth]{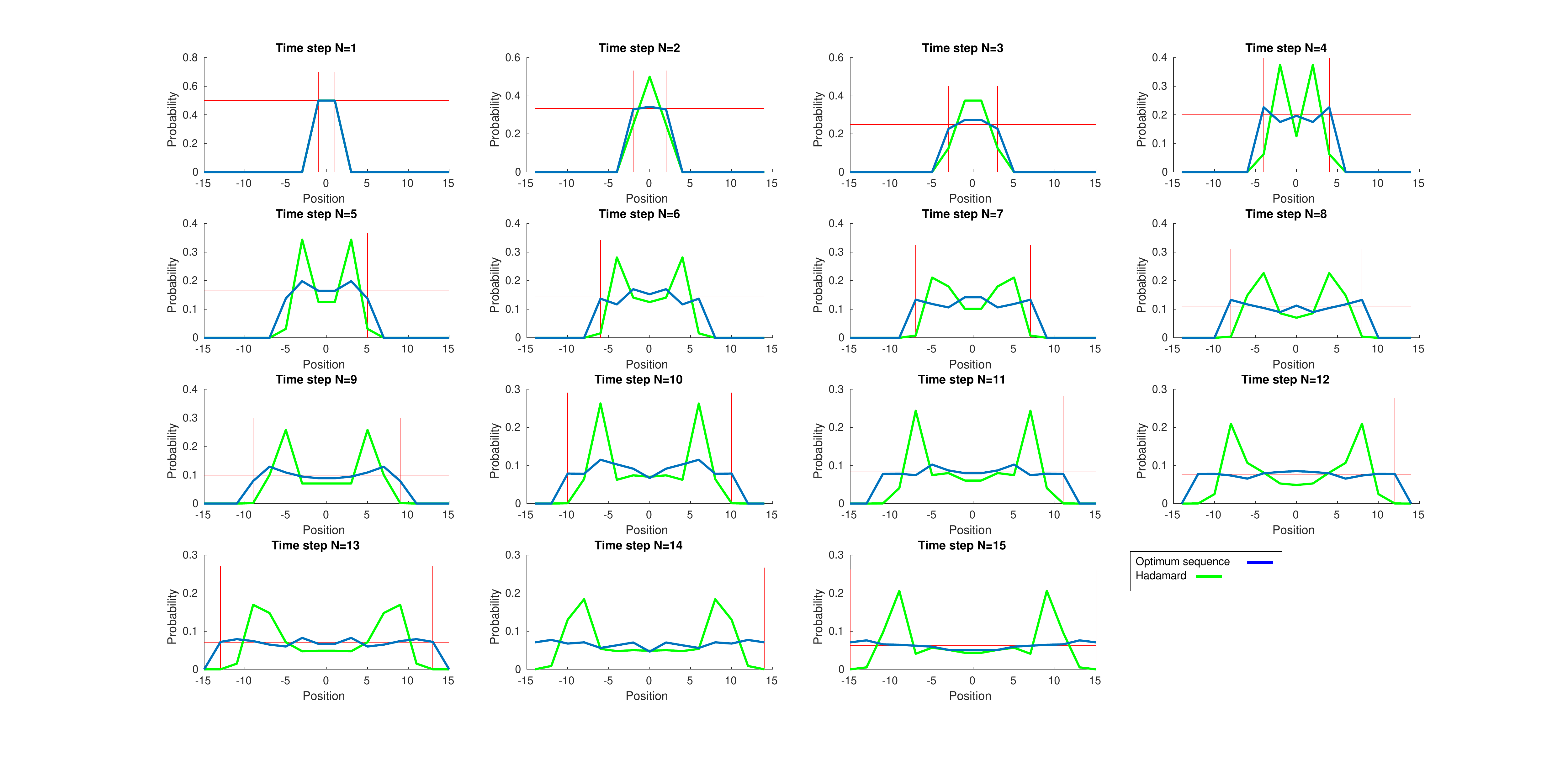}
\caption{\label{fig4} Evolution of the probability distribution in
  position space, $P(x,t)=|\<x|\psi_L(t)\>|^2+|\<x|\psi_R(t)\>|^2$ for
  the optimized set of $\theta$ parameters (blue), and the Hadamard
  coin operator (green) for $T=15$ time steps. The vertical red lines
  represent the maximum possible extension for a quantum walk and the
  horizontal line represents the probability value corresponding to a
  perfectly uniform distribution.}
\end{figure*}

\begin{figure*}
\includegraphics[width=1\textwidth]{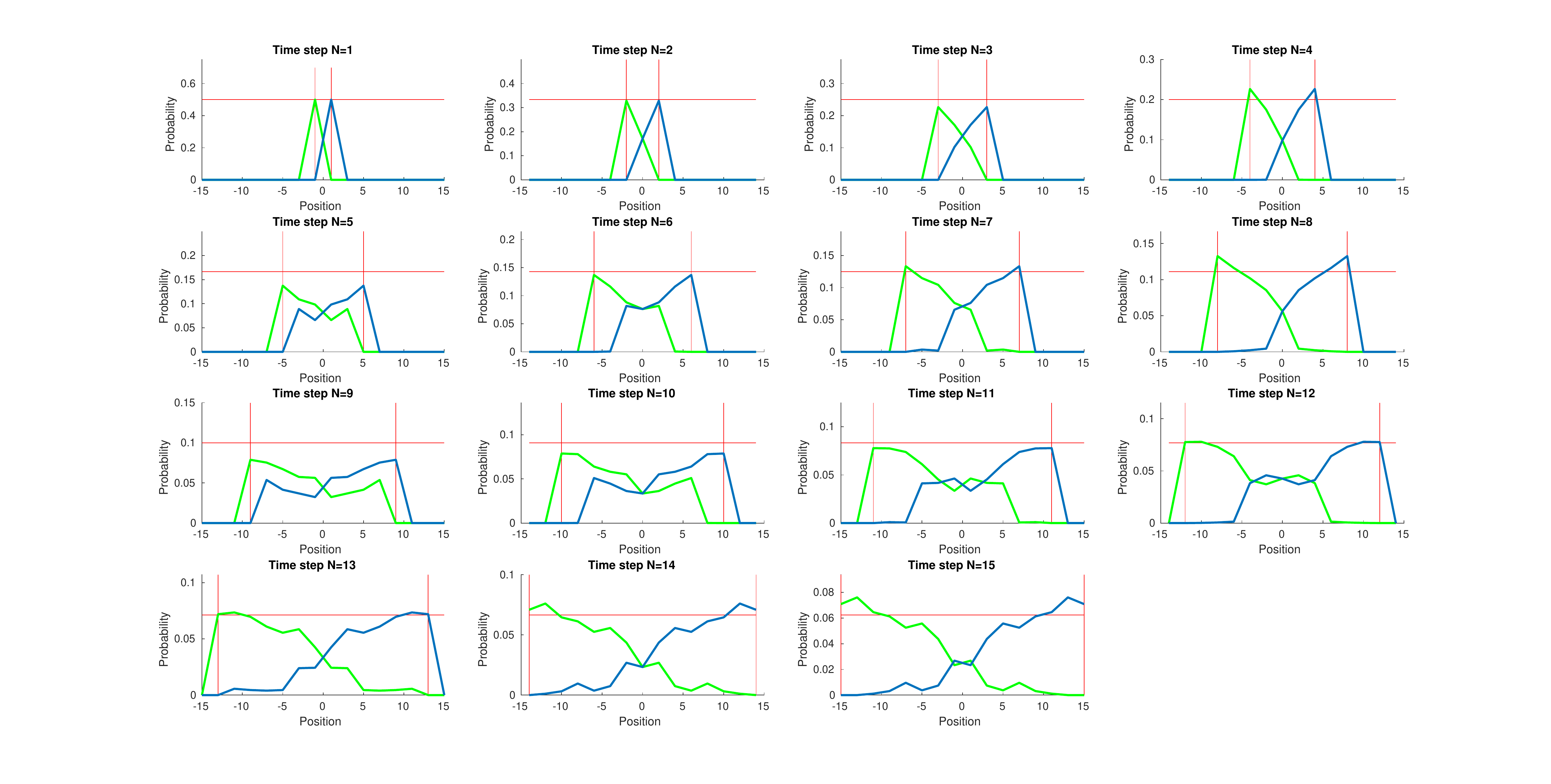}
\caption{\label{fig5} Evolution of the probability distribution in
  position space for $P_R(x,t)=|\<x|\psi_{R}(t)\>|^2$ (blue) and
  $P_L(x,t)=|\<x|\psi_{L}(t)\>|^2$ (green) for the optimized set of
  $\theta$ parameters for $N=15$ time steps. The vertical red lines
  represent the maximum possible extension for a quantum walk and the
  horizontal line represent the probability value corresponding to a
  perfectly uniform distribution.}
\end{figure*}

Let us stress that we require uniformity of the probability in
position space {\em throughout} the $T$ time-steps, not just the last
one. Figure \ref{fig4} illustrates this fact comparing the evolution
of the optimum sequence for $T=15$ with the Hadamard coin operator for
each time step. As we can readily see, the spreading of the optimum
sequence is always larger than in the Hadamard case, and the
probability distribution much more flat. Note that, even though it is
possible to achieve a perfectly uniform distribution up to $T=3$, in
this case there is some deviation because the optimization target is
set to all times up to $T=15$.


\section{Understanding the Optimal Sequences}
\label{sec:understanding}

The optimal values of $\theta_i$ for different final times $T$ are
plotted in Fig. \ref{fig3}G. Although they do not follow a fixed
pattern, the optimal sequences present relevant features, such as a
non-periodic alternation of values $\theta\approx 0$ (coin-tossing
operator close to $\sigma_z$) and $\theta\approx \pi/4$ (close to
$\sigma_x$). In intuitive terms, the values of $\theta\approx 0$ {\em
  split} the wavefunction, making the left and right parts advance
separately in each direction, while values close to $\theta\approx
\pi/4$ {\em combine} both components again. Thus, the optimal
sequences are composed of a certain alternation of both types of
quantum operators: advance and mixture.

\begin{figure*}
\includegraphics[width=1\textwidth]{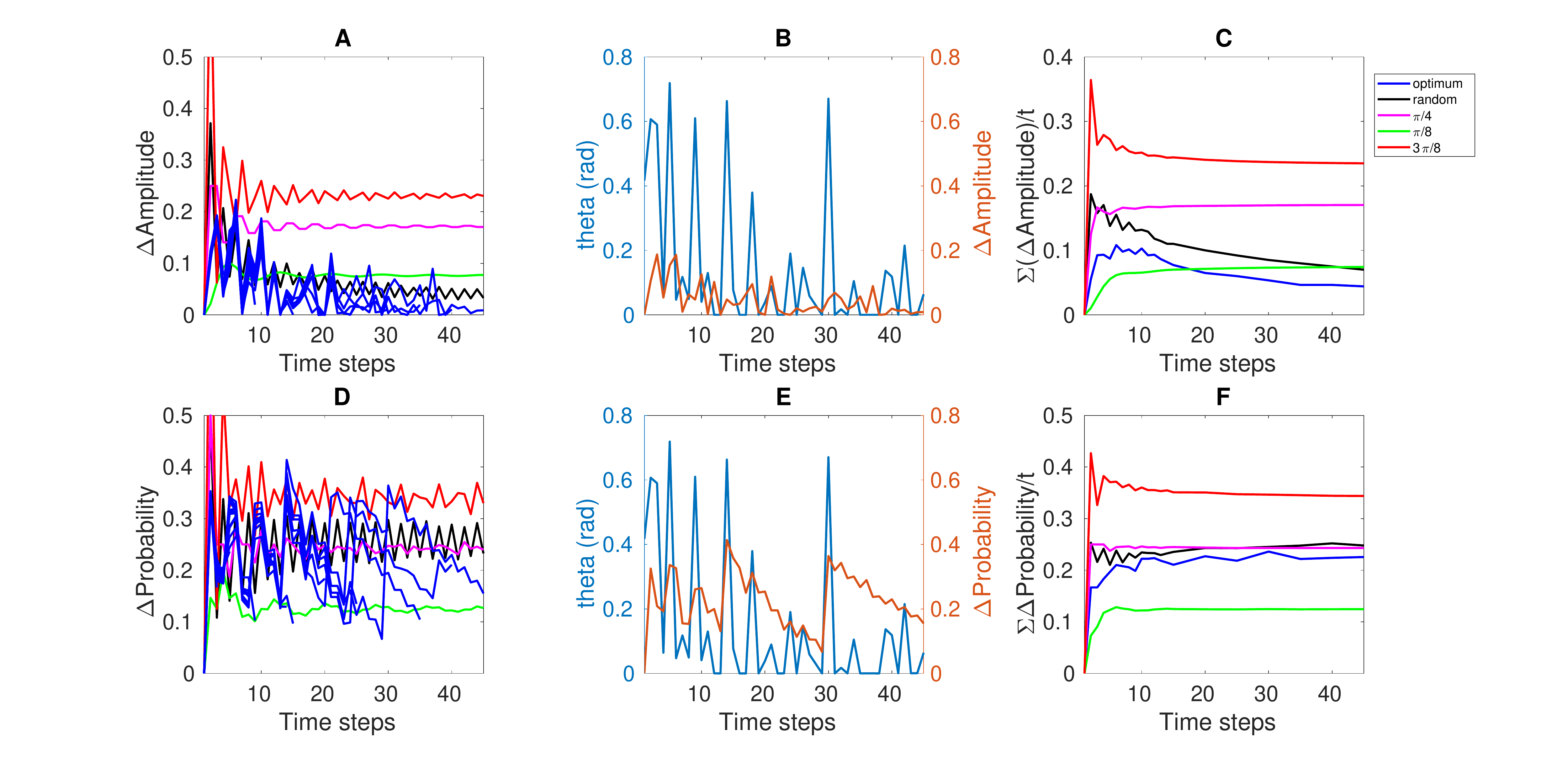}
\caption{\label{fig6} (A) Evolution of the amplitude overlap between
  $\ket{\psi_{R}(t)}$ and $\ket{\psi_{L}(t)}$ for the optimized
  sequences $T=1\cdots 15,20,25,30,35,40,45$. (B) Evolution of the
  amplitude overlap and the $\theta$ parameters for the optimum
  sequence of $T=45$. (C) Evolution of the sum of the normalized
  amplitude overlapping for different optimizations ($T=1\cdots
  15,20,25,30,35,40,45$). (D) Evolution of the probability overlapping
  between $\ket{\psi_{R}(t)}$ and $\ket{\psi_{L}(t)}$ for the
  optimized sequences $T=1..15,20,25,30,35,40,45$. (E) Evolution of
  the probability overlapping and the $\theta$ parameters for the
  optimum sequence of $T=45$. (F) Evolution of the sum of the
  normalized probability overlapping for different optimizations
  ($T=1...15,20,25,30,35,40,45$). The results from the random
  sequences were obtained as the mean value for 500 random
  simulations.}
\end{figure*}

Following Eq. \eqref{eq:psi} we can write the state of the system
$\ket{\psi(t)}$ as

\begin{equation}
  \ket{\psi(t)}=\ket{\psi_L(t)}\ket{L} + \ket{\psi_R(t)}\ket{R}.
  \label{eq:schmidt_dec}
\end{equation}
where $\ket{\psi_{\{L,R\}}}=\sum_x \psi_{x,\{L,R\}}\ket{x}$ need not be normalized
\cite{orthey17}. This allows us to decompose the spatial probability
distribution $P(x,t)=P_L(x,t)+P_R(x,t)$, where
$P_L(x,t)=|\<x|\psi_L(t)\>|^2$ and
$P_R(x,t)=|\<x|\psi_R(t)\>|^2$. Fig. \ref{fig5} shows the time
evolution of both probability distributions. Notice their left-right
symmetry: $P_L(x,t)=P_R(-x,t)$. Moreover, we can also consider the
overlap (or fidelity) between the two wavefunctions:

\begin{equation}
  \Delta_A(t)\equiv \abs{\braket{\psi_R(t)}{\psi_L(t)}}^2,
  \label{over}
\end{equation}
whose behavior is shown in Fig. \ref{fig6}A. We can observe that this
overlap decays towards zero for all values of $T$, faster than for all
other quantum walks, including (an average over) random values. In
Fig. \ref{fig6}B we can see both the overlap and the optimal
$\theta_i$ values for $T=45$, which present little
correlation. Indeed, we can also define an average degree of overlap
along an optimized trajectory,

\begin{equation}
  \Delta_A^{\textrm{norm}}=\frac{1}{T}\sum_{t=1}^T\Delta_t.
  \label{over_norm}
\end{equation}
We show its behavior in Figure \ref{fig6}C. The conclusions are
manifestly disappointing.

Luckily, a slightly different magnitude presents a much more
clarifying behavior. Let us define the {\em probability overlap} as
the area under the minimum:

\begin{equation}
  \Delta_P (t)= \sum_x \min(P_L(x,t),P_R(x,t)),
\end{equation}
which is one if both probability distributions coincide, and zero if
their supports do not intersect. Fig. \ref{fig6}D presents the
time-evolution for the same cases considered in Fig. \ref{fig6}A. In
this case we can observe a {\em sawtooth} behavior in the values of
the probability overlap for the optimal sequence, oscillating around a
finite value. Fig. \ref{fig6}E shows that quick increases in the
probability overlap are caused for large values of $\theta_i \approx
\pi/4$, while small values ($\approx 0$) allow it to decay
linearly. The competition between these {\em pulls} and {\em pushes}
resembles a {\em tug-of-war} which gives rise to the desired optimal
spread. Fig. \ref{fig6}F plots the time-averaged values of the
probability overlap, where we can see that they reach a limit value
which is different from zero. This limit value can be roughly
estimated by considering that the probability distributions $P_L(x,t)$
and $P_R(x,t)$ are approximately linear for long time steps, which can
be verified in the last panels of Fig. \ref{fig5}. In this way, the
probability overlap is the area of an isosceles triangle with altitude
$\frac{1}{2(T+1)}$ and base $N=2T+1$, so for discrete positions and
odd time steps (only odd positions are occupied) we have

\begin{equation}
\begin{split}
A=&\frac{4}{N(N+2)}\sum_{n=1}^{N/2}n \\
 =&\frac{4}{N(N+2)}\sum_{n=0}^{(N-2)/4}(2n+1) \underset{N\rightarrow\infty}{=}\frac{1}{4},
\end{split}
\label{area}
\end{equation}
where we have taken the limit $N \rightarrow \infty$ to obtain our
estimate for the long term probability overlap. Yet, oscillations are
expected for a large time range. Interestingly, the time-averaged
probability overlaps depicted in Fig. \ref{fig6}F for optimal sequeces
are slightly below the (averaged) values obtained for random
sequences, which also tend to a finite value in the long term.

Summarizing, the results obtained so far suggest that the pattern of
optimal $\theta$ parameters is, indeed, complex. Their most salient
feature is a strong alternation of values close to $0$ or to $\pi/4$,
with increasingly long periods. Yet, we show in Appendix
\ref{sec:appendix_D}) that, based on scaling arguments, we can
conjecture that the optimal $\theta$ values will decay in time like
$\propto \arcsin \left( \frac{1}{t} \right)$.

\subsection{Spectral Properties of the Optimal Evolution Operator}

\begin{figure*}
\includegraphics[width=1\textwidth]{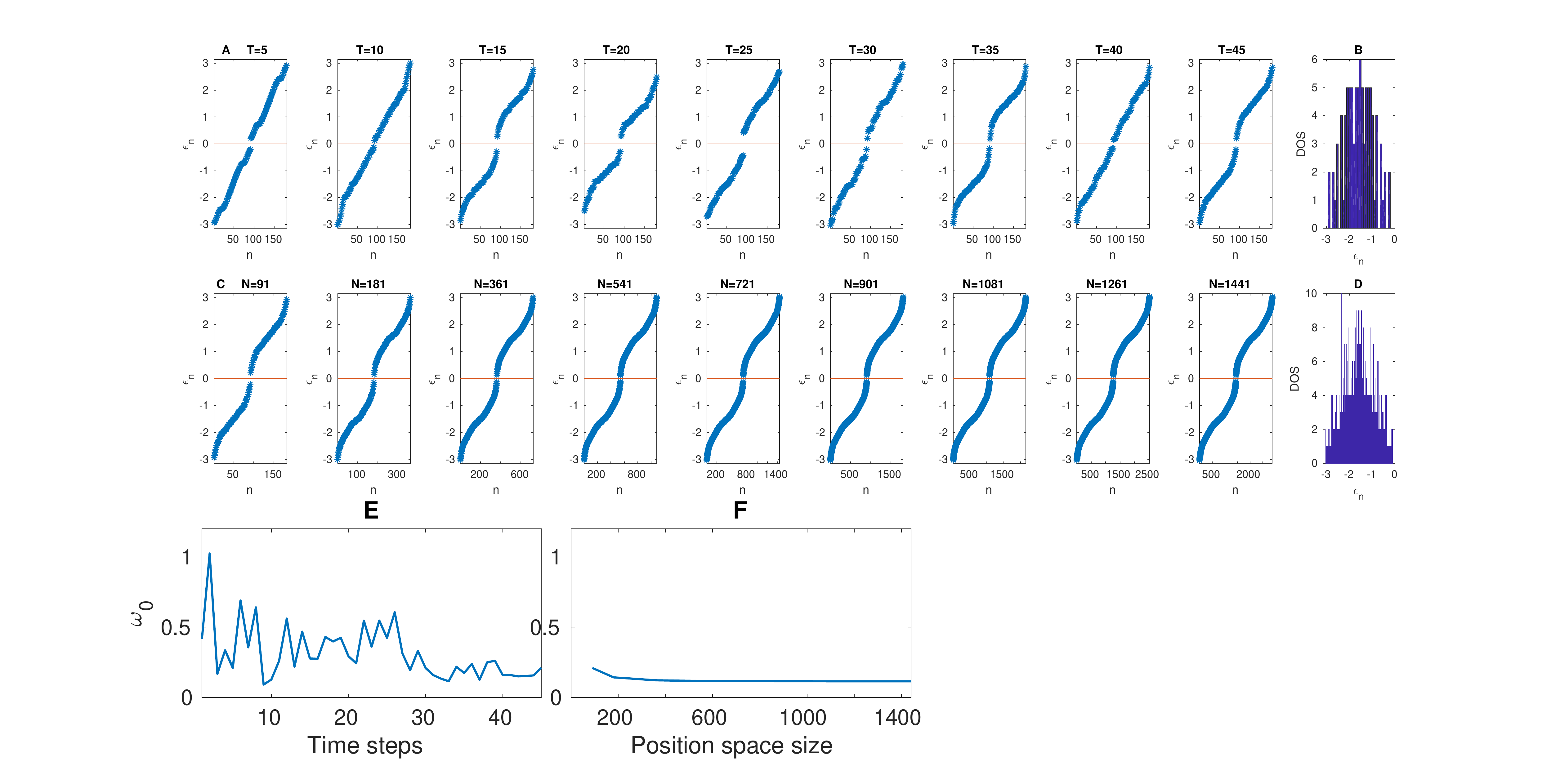}
\caption{\label{fig8} (A) Spectrum $\epsilon_n=i \log(\lambda_n)$
  where $\lambda_n$ are the eigenvalues of the unitary operator
  $\hat{U}(N)$ for different optimization with increasing time steps
  values. (B) Density of states of the lowest band of the spectrum of
  $T=45$. (C) Spectrum $\epsilon_n=i\log(\lambda_n)$ where $\lambda_n$
  are the eigenvalues of the unitary operator $\hat{U}(45)$ with
  increasing positions space size. (D) Density of states of the lowest
  band of the spectrum for the positions space size of $N=1441$. (E)
  Time evolution of the miminum positive value of the spectrum of
  $\hat{U}(45)$ where we have considered that
  $\hat{U}(45)=\hat{U}_{45}...\hat{U}_{1}$. (F) Evolution of the
  minimum positive value of the spectrum of $\hat{U}_{45}$ with
  increasing positions space size.}
\end{figure*}

In this subsection we discuss the energy spectrum of the optimum evolution operator. Since
the evolution is explicitly time-dependent, the system is
non-autonomous and, hence, we cannot define an energy spectrum.  Yet,
we can consider the {\em quasi-energy spectrum}, which can be defined
as $\epsilon_n=i\log(\lambda_n)$ where $\lambda_n$ are the eigenvalues
of the unitary operator $\hat{U}_N$ \cite{gullo17, buarque19}. It is
interesting to consider the asymptotic properties of the system for
which it is necessary to study the spectrum of the total evolution
operator for a time step $t$ such
$\hat{U}(t)=\hat{U}_t\hat{U}_{t-1}...\hat{U}_2\hat{U}_{1}$, where
$t\gg 1$. Due to the computational cost of optimizing quantum walks
for large time lapses, we are limited to $T=50$. It has been shown, at
least for certain aperiodic sequences as well as for simple periodic
ones, that after few steps ($t\sim 30$) the spectrum of the total
evolution operator does not change appreciably \cite{gullo17}.

We show the results in Fig. \ref{fig8}. First, we obtain the spectra
for different optimizations corresponding to increasing time steps
(Fig. \ref{fig8}A,B), where the size of the position space is the
minimum possible to avoid boundaries (i.e. $2T+1$). As we are
considering finite-dimension Hilbert spaces, we obtain discrete point
spectra, where the appearence of a gap around $\epsilon_k=0$ can be
appreciated. It is interesting to compare this quasi-energy spectrum
with the asymptotic spectra of certain aperiodic and periodic
sequences, where such a gap does not appear \cite{gullo17}. Let us
consider the spectrum of $\hat{U}(T=45)$ for different system sizes,
where we can see that the gap is maintained in Fig. \ref{fig8}C,D.

We can see also consider the time evolution of the quasi-spectral gap
of $\hat{U}(t)$, fixing $T=45$, as it is shown in
Fig. \ref{fig8}E. Notice that, despite the fluctuations, it seems to
tend to a finite value, although larger time lapses would be required
in order to confirm this tendency. The quasi-spectral gap does not
possess a relevant dependence on the system size $N$, as we can see in
Fig. \ref{fig8}F.

The energy spectrum for constant $\theta $ values can be interpreted
as a dispersion relation $E(k)$ \cite{buarque19}. As the quasi-energy
spectrum remains unchanged for long enough times (Fig. \ref{fig8}A,
\citep{gullo17}) it is natural also to interpret it as a dispersion
relation. Furthermore, for constant values of $\theta$ we can obtain
an analogue of the Klein-Gordon equation for $\psi_R$ and $\psi_L$,
where the mass is given by \cite{chandra10}

\begin{equation}
  M=\sqrt{2(\sec{(\theta)}-1)\over\cos{(\theta)}},
\end{equation}
This implies that for $\theta=0$ the quantum walker is equivalent to a
massless particle, presenting a gapless linear spectrum and non-zero
group velocity. On the other hand, for increasing values of $\theta$
there appears an increasing gap (and therefore an increasing mass)
\cite{buarque19}. For $\theta=\pi/2$ we have an particle with infinite mass and
zero group velocity (its spectrum is flat and gapped). Moreover, it
has been shown that periodic and some aperiodic sequences yield
gapless spectra, which can be both linear and non linear, respectively
\cite{gullo17}. They can be understood to represent massless particles of
constant and variable group velocity, respectively.

Our optimal sequences present a non linear gapped spectrum, similar to
the random sequences, so they can be understood as the evolution of a
massive particle with variable group velocity. The reason can be
described as follows. In order to explore space efficiently, the
quantum walker should be able to reach the maximal possible spread
with a finite probability, but it should also reach all other possible
sites, with a similar probability. Thus, it should evolve with
different propagation velocities, ranging from the maximal velocity,
corresponding to $\theta=0$ and the minimal one, corresponding to
$\theta=\pi/2$. The group velocities are evaluated from

\begin{equation}
  v_g(k)={dE(k)\over dk},
\end{equation}
which covers a broad range, as we can see from Fig.\ref{fig8}.

\subsection{Survival probability} 

In order to understand the asymptotic dynamics of the system (and study its behavior in relation with the spectral properties) we introduce the survival probability, which
it is defined via its amplitude

\begin{equation}
\nu(t)=\braket{\psi(0)}{\psi(t)},
\label{surv}
\end{equation}
where $\ket{\psi(t)}=\hat{U}(t)\ket{\psi(0)}$. Physically, it
describes the probability of finding the state in the time step $t$ in
the initial state, that in our case is
$\ket{\psi(0)}=\frac{1}{\sqrt{2}}\ket{0}(\ket{R}+i\ket{L})$. Note that
it is not strictly the probability of finding the evolved state in the
initial position, but in the initial state. The survival amplitude is
also directly related to the Fourier transform of the spectral measure
of the evolution operator \cite{last96} so that

\begin{equation}
  \abs{\nu(t)}^2 =\abs{\int_\sigma{}d \mu_0 (\epsilon) e^{-i\epsilon t}}^2,
\label{survfour}
\end{equation}
where $ \mu_0 $ is the measure induced by the initial state. We have
the important result that the Fourier transform of the survival
probability is the measure itself.

\begin{figure*}
\includegraphics[width=1\textwidth]{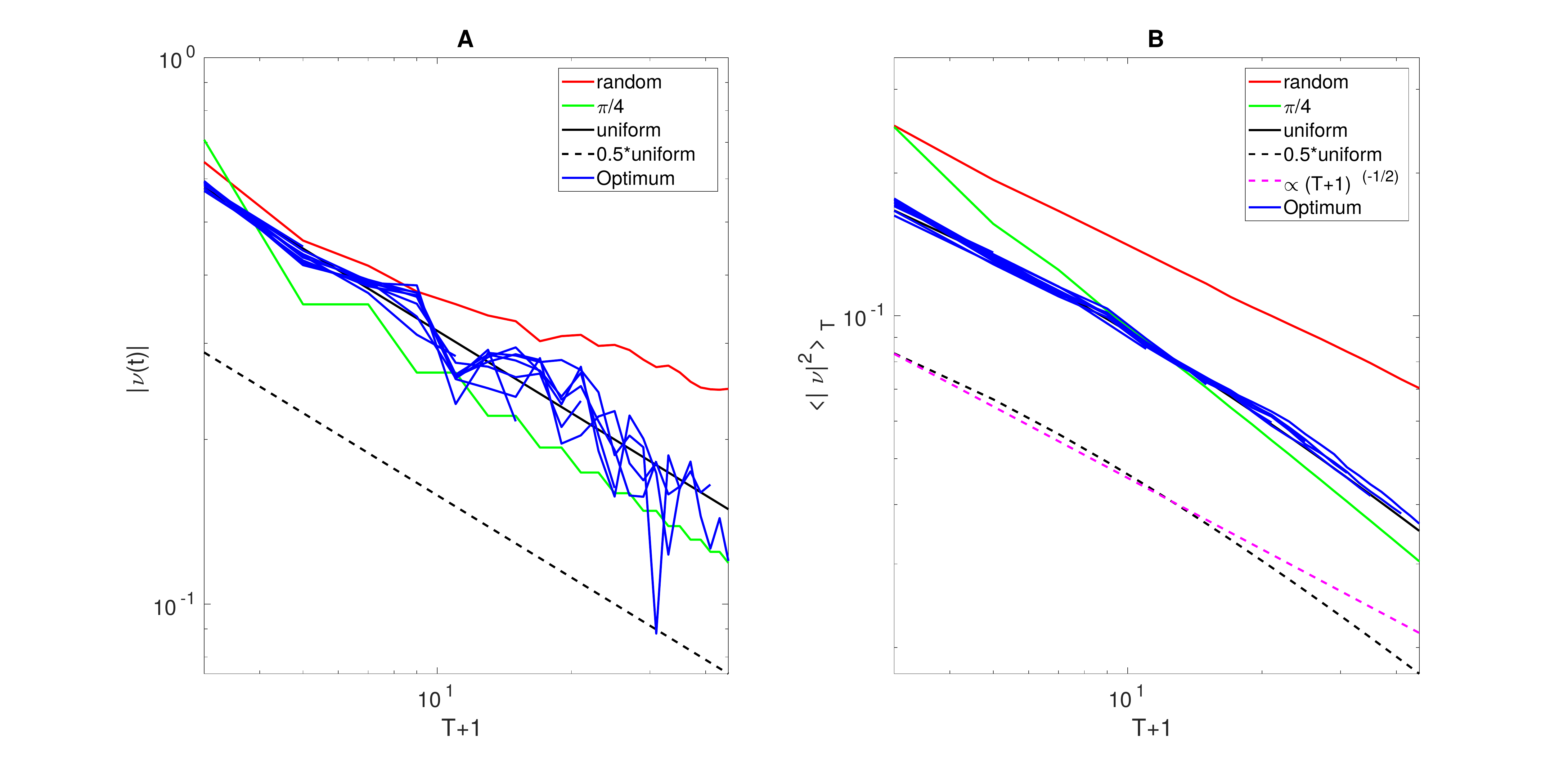}
\caption{\label{fig7} (A) Absolute value of the survival amplitude
  $\abs{\nu(t)}$ for different optimum sequences obtained for
  differents time steps ($T=1...15,20,25,30,35,40,45$). (B)
  Corresponding Ces\'aro averages of $\abs{\nu(t)}^2$. In both cases,
  the black line corresponds to the maximum values of a perfectly
  uniform distribution (\eqref{survuni} in A and \eqref{cesarouni} in B) and the red line to the mean value of $N_s=500$ random simulations. We have added the uniform distribution with a factor of 0.5 (black dashed line) for a better understanding due to the complete overlapping of the optimum sequences (blue lines) with the uniform distribution (black line). For the Ces\'aro averages we have added a $ 1/\sqrt{T+1}$ function (magenta dashed line) for comparative purposes; the uniform distribution corresponds to \eqref{cesarouni}.}
\end{figure*}

We will also obtain the time average of the survival
probability (Ces\'aro average) defined as

\begin{equation}
\ \langle \abs{\nu}^2 \rangle_T=\frac{1}{T}\sum_{t=1}^T\abs{\nu(t)}^2.
\label{cesaro}
\end{equation}
We can obtain the maximum survival probability and its Ces\'aro
average for an idealized uniform distribution in position space given
as

\begin{align}
\ \nu(t)^{uni}&=\frac{1}{\sqrt{t+1}},
\label{survuni} \\
\ \langle \abs{\nu}^2 \rangle_T^{uni}&=\frac{1}{T}\sum_{t=1}^T\frac{1}{t+1}.
\label{cesarouni}
\end{align}
We represent in Figure \ref{fig7}A the absolute value of the survival
probability, $|\nu(t)|$, computed with \eqref{survuni}, for different
quantum walks, as in the previous sections. Concretely, we use
optimized sequences for several values of $T$, (averaged) values for
random sequences and the $\theta_i=\pi/4$ quantum walk. Moreover, we
also compare to the {\em uniform} wavefunction, which is given by

\begin{equation}
  \ket{\psi_U(T)}={1\over\sqrt 2} \sum_{x=-T}^T \(\ket{x,L}+\ket{x,R}\).
\end{equation}
The right panel, \ref{fig7}B, shows the Cesáro averaged
values. Clearly, random sequences provide the largest value (in
average) for the survival probability, while optimal and uniform
values stay between the random and the Hadamard cases. Indeed, the
uniform and the optimal values remain similar for all times. The behaviour of the Ces\'aro averages is quite similar.

There are important connections between the survival probability and
its Ces\'aro average and the spectral properties of the system. The
spectral measure can be splitted into three parts: pure point,
singular continuous and absolute continuous \cite{fillman17}. Pure
point spectrum is usual for disorderd systems when the $ \theta $ is
randomly distributed while the absolutely continuous spectrum is
related to highly structured systems, such as periodic sequences of
quantum coins \cite{fillman17}. The singular continuous spectrum
appears in between, for example when there is aperiodicity
\cite{gullo17}. Heuristically, we can assert that the amount of order
in the coin parameters is directly related to how continuous the
spectral measure is.

For long times, we use the fact that the survival amplitude is the
Fourier transform of the measure \eqref{survfour}, so we can extract
information of the spectrum studying the long-term behaviour of
\eqref{surv} and \eqref{cesaro}. Specifically, we use the conditions
derived from Wiener's lemma \cite{queffelec10} and the theorem of
Ruelle, Amrein-Georgescu and Enss (RAGE) (\cite{last96};
\cite{fillman17} for the discrete-time version). Indeed, the following
conditions

\begin{align}
\lim_{t\to\infty} \< \abs{\nu}^2 \>_T=0,
\label{cesaroinf} \\
\lim_{t\to\infty} \abs{\nu(t)}=0,
\label{survinf} 
\end{align}
imply that the spectrum of the evolution operator will be absolutely
continuous. The first one, Eq. \eqref{cesaroinf}, guarantees that the
spectrum lacks a pure point part and the second one,
Eq. \eqref{survinf}, ensures that the spectrum is absolutely
continuous. Despite our computational limitations regarding the
maximal time-step, we can extract relevant information from the
time-evolution of the idealized uniform system, since we know that the
survival probability and Ces\'aro averages are similar to those of the
optimized sequences. This implies that, since both conditions
\eqref{cesaroinf} and \eqref{survinf} are met, the spectrum of the
uniform quantum walk in the limit $t\rightarrow\infty$ is absolutely
continuous. It is interesting to note that the aperiodic sequences
commented above induce a singular continuous energy spectrum since
\eqref{survinf} is not met, but the periodic sequences behave
similarly as the optimum sequence since both conditions are met
yielding absolutely continuous energy spectra \cite{gullo17}.

As noted in \cite{fillman17}, using the discrete-time version of the RAGE theorem we can relate the different spectral types commented above with the localization/spreading behavior of the wavefunction:

\begin{itemize}
\item Pure point: most of the wavepacket never leaves a given bounded
  region, so the wavefunction will remain localized.

\item Singular continuous: upon time-averaging, the wavepacket will
  eventually leave any bounded region, but this could not be true of
  all walks.

\item Absolutely continuous: most of the wavepacket will eventually
  leave any bounded region. In one dimension the spread will be
  ballistic upon time averaging and, without averaging, for some
  specific quantum walks.
\end{itemize}

Asymptotically, the behaviour of the optimum sequence is expected to
be similar to that studied for finite times: the wavepacket spreads
ballistically leaving any bounded region. Curiously, this is also the
behaviour of periodic sequences: the optimal sequence seems not be
periodic, but asymptotically behaves as a highly ordered periodic
sequence. Contrarily, the aperiodic sequences can show {\em anomalous
  transport}, i.e. the wavepacket leaves any bounded region with
sub-ballistic speed. Furthermore, we have the expected analytical
expression for the survival probability \eqref{survuni} and the
Ces\'aro average \eqref{cesarouni} of the optimum sequence, providing
the exact exponent of the power-law decay; for aperiodic sequences the
maximum value of this exponent is $ \sim0.8 $ \cite{gullo17}. It is
also interesting to note that random sequences do not seem to fulfill
conditions Eq. \eqref{cesaroinf} and Eq. \eqref{survinf}, implying
that they present a pure point spectrum yielding a localized
wavefunction.


\section{Conclusions}
\label{sec:conclusions}

In this article we have provided a protocol to build a time-dependent
quantum walk which provides {\em optimal spatial spread}, i.e.: for
which the spatial distribution is (nearly) maximal for all time
steps. We have restricted ourselves to coin-tossing operators with a
single parameter, $\theta$, because the inclusion of all three Euler
angles does not improve the results substantially. The optimal
sequences depend on the maximal time considered, $T$, and present
complex structures. Yet, some patterns arise. First of all, most
values are close to either $\theta=0$ or $\theta=\pi/4$. The first
values tend to {\em stretch} the left and right parts of the
wavefunction. The second ones tend to appear when the {\em probability
  overlap} has fallen below a certain threshold, and allow the
wavefunction to combine again. Even though finding the optimal
sequences can be a complicated optimization problem, a crude estimate
can be obtained.

We have considered the long-term dynamics associated with these
optimal quantum walk, regarding the survival probability (the fidelity
with the initial state) and the spectrum of the evolution
operator. The observed behavior leads to conjecture that the spectrum
is absolutely continuous, a behaviour typical of highly ordered sequences as the periodic ones.

Regarding lines of future work, we are interested in {\em
  quasi-optimal sequences}, alternating $\theta=0$ and $\theta=\pi/4$
values in a regular (although non-trivial) pattern which will give
rise to a quasi-optimal spread. Indeed, these quasi-optimal sequences
will be much easier to obtain in the laboratory. Moreover, we intend
to obtain the corresponding values in $>1D$ and disordered
lattices. It is very relevant to consider the search capabilities of
these optimized quantum walks, which will be substantially improved
over time-independent or random quantum walks.

\section{Acknowledgements}

We would like to acknowledge Silvia N. Santalla and Germán Sierra for
useful discussions. This work has been partially funded by the Spanish
Government and the European Union through grant QUITEMAD-CM
P2018/TCS-4342. 

\newpage


\appendix

\section{Boundary conditions for the evolution operator}
\label{sec:appendix_A}

As mentioned in section \ref{sec:model}, the unitary evolution
operator \eqref{evolution} is defined on an infinite dimensional
Hilbert space associated to all possible positions. For practical
purposes, (i.e. numerical analysis) we need to consider a finite
dimensional positions space, but this may turn non-unitary both the
shift operator \eqref{shift} and, therefore, the evolution operator
\eqref{evolution}. We solve this issue, formally, by defining cyclic
boundary conditions in such a way that the shift operator is defined
by $\hat{S}=\hat{S}_c+\hat{S}_b $ where

\begin{equation}
\begin{split}
\hat{S}_c=&\left(\sum_{x=1}^{N-1}\ket{x+1}\bra{x}\right)\otimes\ket{R}\bra{R}+\\
&\quad      +\left(\sum_{x=2}^{N}\ket{x-1}\bra{x}\right)\otimes\ket{L}\bra{L},
\end{split}
\label{shiftc}
\end{equation}

\begin{equation}
\hat{S}_b=\ket{1}\bra{N}\otimes\ket{R}\bra{R}+\ket{N}\bra{1}\otimes\ket{L}\bra{L},
\label{shiftc}
\end{equation}
where $N$ is the finite size of the position space. We always consider
that the particle starts in the middle of the positions space (i.e. $
x=0$), so if we set the size of the positions space as $N=2T+1$, where
$T$ is the number of time steps, the particle never actually
experiences the boundary conditions. Nevertheless, when analyzing
spectral properties of the evolution operator the boundary conditions
are evaluated. 


\section{QRW fails to be uniformly distributed in space for $N=4$}
\label{sec:appendix_B}

We now show that a QRW using a time-dependent coin operator of the
form \eqref{coinop2} can not be uniformly distributed in positions
space when considering a symmetric initial state \eqref{initial}.

Considering that the probability for the particle to be at $x=i$ at
time-step $t$ is given by

\begin{equation}
  Pr_i(t)=\Tr[( \ket{i}\bra{i} \otimes \hat{\mathbb{I}}_p) \cdot
    \hat{\rho}(t)],
\end{equation}
we have that, for all $t$,

\begin{equation}
\sum_{x=-t}^{t} \Tr[\left( \ket{x}\bra{x} \otimes \hat{\mathbb{I}}_p
  \right) \cdot \hat{\rho}(t)]=1,
\label{prob}
\end{equation}
where $\hat{\rho}(t)=\ket{\psi(t)}\bra{\psi(t)}$ is the density
operator at time step $t$, and

\begin{equation}
\hat{\rho}(t)=\hat{U}_t\cdots\hat{U}_1\hat{\rho}(0)
\hat{U}_1^{\dag}...\hat{U}_t^{\dag}.
\end{equation}
This is valid for any initial state $\hat{\rho}(0)$, but from here we
will consider that $\hat{\rho}(0)=\ket{\psi_S}\bra{\psi_S}$, where
$\ket{\psi}_S$ is the symmetric state \eqref{initial}. Note that, due
to the cyclic properties of the trace, for an uniform probability
distribution in positions space we should have

\begin{equation}
\ \Tr[\hat{P}(t) \cdot \hat{\rho}(0)]=\frac{1}{t+1}
\label{probuni}
\end{equation}
where $\hat{P}(t)=
\hat{U}_1^{\dag}...\hat{U}_t^{\dag}\left(\ket{0}\bra{0} \otimes
\hat{\mathbb{I}}_c\right)\hat{U}_t...\hat{U}_1$ for even time steps
and $\hat{P}(t)=
\hat{U}_1^{\dag}...\hat{U}_t^{\dag}\left(\ket{1}\bra{1} \otimes
\hat{\mathbb{I}}_c\right)\hat{U}_t...\hat{U}_1$ for odd time
steps. Thus, it would be necessary only to evaluate the operator
$\hat{P}(t)$.

Let us prove it directly by evaluating \eqref{prob} for $t=1,2,3,4$
and substituting the solutions sequentially, searching for the values
of $\{\theta_i\}$ that will make all the spatial probabilities equal.

\begin{itemize}
\item $t=1$

\begin{equation}
\ Pr_{x=-1}=Pr_{x=1}=\frac{1}{2},
\label{prob1}
\end{equation}
and there is no dependence on $\theta_1$. Thus, $\theta_1$ can take
any value in $[0,\frac{\pi}{2}]$.

\item $t=2$

\begin{equation}
\begin{split}
\ Pr_{x=-2}&=Pr_{x=2}=\frac{1}{2}\cos^2(\theta_2), \\
\ Pr_{x=0}&=\sin^2(\theta_2),
\end{split}
\label{prob2}
\end{equation}
whose solution is $\theta_2=\arctan(\frac{1}{\sqrt{2}})$.

\item $t=3$
\begin{equation}
\begin{split}
\ Pr_{x=-3}=Pr_{x=3}&=\frac{1}{3}\cos^2(\theta_3), \\
\ Pr_{x=-1}=Pr_{x=1}&=\frac{1}{2}-\frac{1}{3}\cos^2(\theta_3),
\end{split}
\label{prob3}
\end{equation}
whose solution is $\theta_3=\frac{\pi}{6}$.

\item $t=4$
\begin{equation}
\begin{split}
\ Pr_{x=-4}&=Pr_{x=4}=\frac{1}{4}\cos^2(\theta_4), \\
\ Pr_{x=-2}&=Pr_{x=2}=\frac{1}{3}-\frac{1}{6}\cos^2(\theta_4)+\frac{\sqrt{2}\cos(\theta_4)\sin(\theta_4)}{12}, \\
\ Pr_{x=0}&=\frac{1}{3}-\frac{1}{6}\cos^2(\theta_4)-\frac{\sqrt{2}\cos(\theta_4)\sin(\theta_4)}{12},
\end{split}
\label{prob4}
\end{equation}
and, in this case, the system is incompatible.
\end{itemize}


\section{Optimization for the final step}
\label{sec:appendix_C}

\begin{figure*}
\includegraphics[width=1\textwidth]{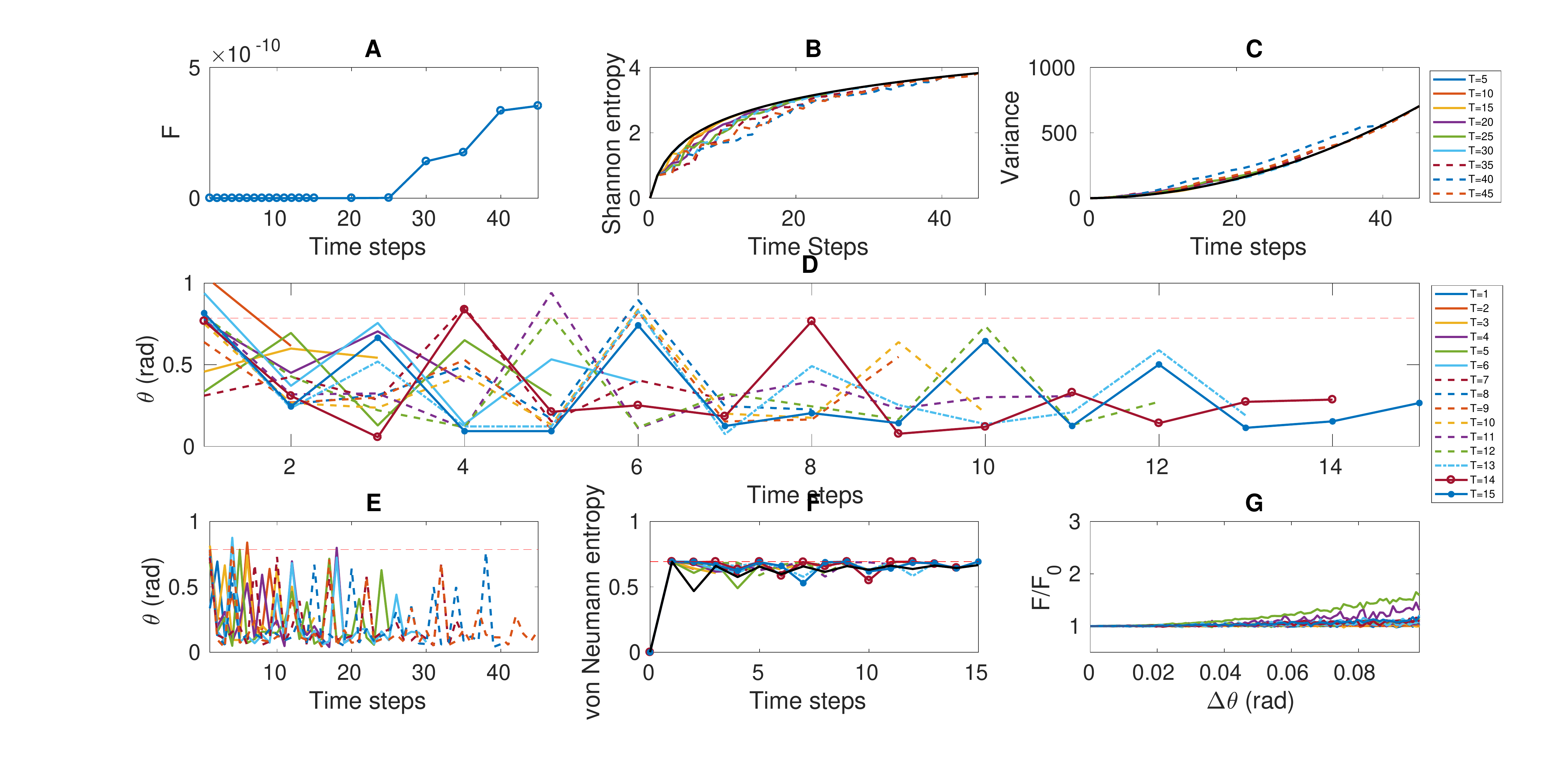}
\caption{\label{figA} Optimized sequences of the $\theta$ sequences
  for $T=1...15,20,25,30,35,40,45$ for the last-time step
  optimization, Eq. \eqref{F'}. (A) Evolution of the $F'$ value
  for different time steps optimizations. (B) Exponential of
  the Shannon entropy of the probability distribution in positions
  space and (C) variance for the optimum sequence for
  $T=5,10,15,20,25,30,35,40,45$. (D) Optimized sequences of
  $\theta $ parameters for $T=1$ up to $T=15$. The horizontal dashed
  red line corresponds to $\frac{\pi}{4}$ (Hadamard). (E)
  Optimized sequences of $\theta$ parameters for
  $T=1..15,20,25,30,30,35,40,45$. The horizontal dashed red line
  corresponds to $\frac{\pi}{4}$ (Hadamard). (F) Time
  evolution of the von Neumman Entropy of the reduced density matrix
  as a measure of entanglement for the optimized sequences of $\theta$
  parameters for $T=1$ to $T=15$. The black line corresponds to the
  mean value of 1000 simulations with random $\theta$ parameters. The
  horizontal red dashed line corresponds to the situation of maximum
  entanglement $\log(2)$. (G) Stability of the optimized
  sequences of $\theta$ parameters against increasing perturbations
  (noise) of the $\theta$ value (see text).}
\end{figure*}

As we did in Figure \ref{fig3}, let us perform the optimization
process imposing the only condition that the probability distribution
be uniform in the last step, where instead of using Eq. \eqref{F} we
define a new function

\begin{equation}
F'(\theta_{0},...,\theta_{T})=1-\frac{S(\theta_{0},...,\theta_{T})}{S_{max}^{T}},
\label{F'}
\end{equation}
where $S$ is the Shannon entropy. Function \eqref{F'} is normalized so
that $F=1$ for a completely localized particle, and $F=0$ for a
completely uniform distributed particle in the last time step. The
results are shown in Fig. \ref{figA}. As expected, the values for $F$
are considerably lower (seven orders of magnitude) and the Shannon
entropy only matches the maximum value at the last time
step. Regarding the $\theta$ values, they exceed $\frac{\pi}{4}$ and
do not approach zero. Moreover, they do not show any recognizable
pattern. As expected, the robustness is high and the procedure
tolerates higher noise in the parameters. The optimization in the last
step could be useful, for example, for preparing the system in a
uniform distribution for a given time step.

\section{Optimal parameters in the long time limit}
\label{sec:appendix_D}

Let us estimate the behavior of the optimal $\theta$ values for long
times. Lets consider the positions $x=+1$ and $x=-1$ for an odd time
step $t$ and $x=0$ for even time steps $t+1$ so that

\def\up{\uparrow}
\def\dn{\downarrow}

\begin{align}
  \ &\ket{\psi_{-1}(t)}=\ket{-1}
  \( r_{-1}\ket{\up}+l_{-1}\ket{\dn} \),
\label{psi_1} \\
\ &\ket{\psi_{1}(t)}=\ket{1} \( r_{1}\ket{\up}+l_{1}\ket{\dn} \),
\label{psi1} \\
\ &\ket{\psi_{0}(t+1)}=\ket{1} \( r_{0}\ket{\up}+l_{0}\ket{\dn} \),
\label{psi0}
\end{align}
where $r_{-1}$, $r_1$, $l_{-1}$, $l_1 \in \C$ are probability
amplitudes (where we have omitted the time dependence), and $r_0$ and
$l_0$ are given by

\begin{align} 
\ &r_0=r_{-1} \cos\theta_t + l_{-1}\sin \theta_t,
\label{r0} \\
\ &l_0=r_{1} \sin\theta_t - l_{1}\cos \theta_t.
\label{l0}
\end{align}
Due to the uniform probability distribution restriction and that
$|\psi_R(t)|^2$ and $|\psi_L(t)|^2$ are symmetric with respect to
$x=0$, the following conditions are fulfilled

\begin{align}
\ &|r_{-1}|^2+|l_{-1}|^2=|r_{1}|^2+|l_{1}|^2=\frac{1}{t+1},
\label{cond1} \\
\ &|r_{1}|^2=|l_{-1}|^2,
\label{cond2} \\
\ &|r_{-1}|^2=|l_{1}|^2,
\label{cond3} \\
\ &|r_{0}|^2=|l_{0}|^2=\frac{1}{2(t+2)}.
\label{cond4}
\end{align}
Considering \eqref{cond2}, \eqref{cond3} and \eqref{cond4} we get

\begin{equation}
r_1^*l_1+r_1l_1^*=-(r_{-1}^*l_{-1}+r_{-1}l_{-1}^*),
\label{cond5}
\end{equation}
that can be expressed as

\begin{equation}
\Re(r_1^*l_1)=-\Re(r_{-1}^*l_{-1}),
\label{cond5b}
\end{equation}
so that condition $|r_{0}|^2+|l_{0}|^2=\frac{1}{t+2} $ results in

\begin{equation}
  |r_1|^2 \sin^2\theta_t+|r_{-1}|^2 \cos^2\theta_t-
  2\cos\theta_t\sin\theta_t\Re(r_1^*l_1)=\frac{1}{2(t+2)}.
\label{thetaevol}
\end{equation}
Let us make a further assumption: $|r_1|^2\approx|r_{-1}|^2 $ for
$t\gg 1$, if we can assume that probability distributions are linear
or, at least, smooth enough, as the numerical results suggest. Thus,
\eqref{thetaevol} turns into

\begin{equation}
  4\cos\theta_t\sin\theta_t
  \Re(r_1^*l_1)\sim \frac{1}{t+1}-\frac{1}{t+2}\sim\frac{1}{t^2}.
\label{thetaevol2}
\end{equation}
Considering condition \eqref{cond1} it can be expected that
$\Re(r_1^*l_1)\sim\frac{1}{t}$ and therefore $\theta_t\sim\frac{1}{2}
\arcsin\(\frac{8}{t} \)$. So, for long times it is expected that the
values of $\theta$ to be slowly decaying.


\end{document}